\pdfoutput=1
\documentclass[aps,pra,preprint,tightenlines,showpacs,showkeys,superscriptaddress]{revtex4-1}
\usepackage{graphicx}
\usepackage{amssymb}
\usepackage[squaren, cdot]{SIunits}
\usepackage[latin9]{inputenc}

\begin{document}
\title{Scalable Architecture for Quantum Information Processing with Atoms in Optical Micro-Structures}
\author{Malte Schlosser}
\author{Sascha Tichelmann}
\author{Jens Kruse}
\author{Gerhard Birkl}
\email[]{gerhard.birkl@physik.tu-darmstadt.de}
\affiliation{Institut f\"{u}r Angewandte Physik, Technische Universit\"{a}t Darmstadt, Schlossgartenstra\ss e 7, 64289 Darmstadt, Germany}
\date{\today}
\newcommand{\bra}[1]{\ensuremath{\langle #1 |}}
\newcommand{\ket}[1]{\ensuremath{| #1 \rangle}}
\newcommand{\braket}[2]{\ensuremath{\langle #1 | #2 \rangle}}

\begin{abstract}
We review recent experimental progress towards quantum information processing and quantum simulation using neutral atoms in two-\-di\-men\-sio\-nal (2D) arrays of optical microtraps as 2D registers of qubits. We describe a scalable quantum information architecture based on micro-fabricated optical elements, simultaneously targeting the important issues of single-site addressability and scalability. This approach provides flexible and integrable configurations for quantum state storage, manipulation, and retrieval. We present recent experimental results on the initialization and coherent one-qubit rotation of up to 100 individually addressable qubits, the coherent transport of atomic quantum states in a scalable quantum shift register, and discuss the feasibility of two-qubit gates in 2D microtrap arrays.
\end{abstract}
\maketitle


\section{Introduction}
\label{sec:intro}
%
\begin{figure*}
\begin{center}
 \includegraphics[width=0.75\textwidth]{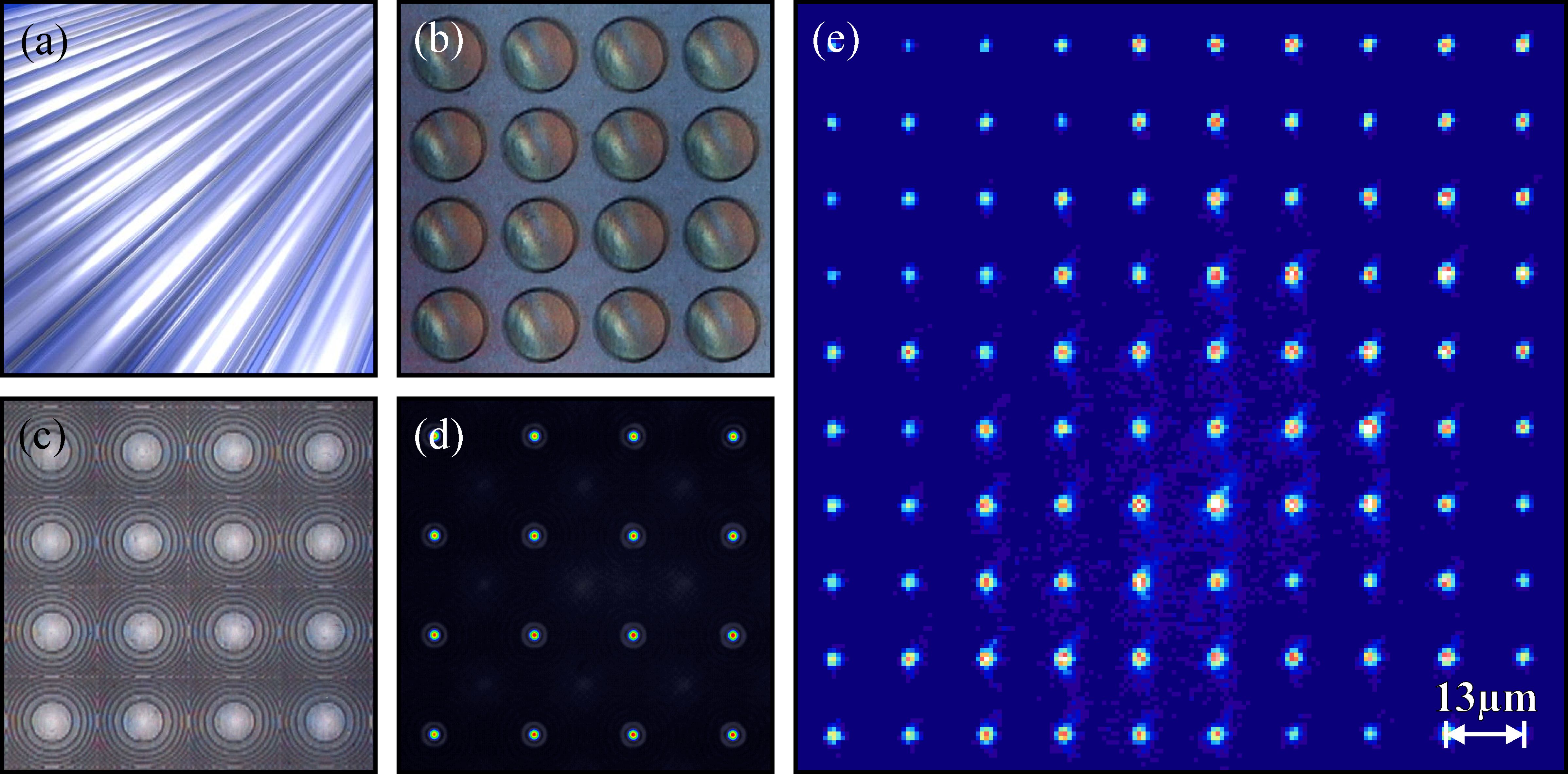}
 \end{center}
\caption{
Typical examples of micro-fabricated optical elements, the generated light fields, and a fluorescence image of atoms trapped in the resulting potential geometry: (a) 1D array of refractive cylindrical microlenses (pitch: $\unit{400}{\micro\meter}$); 2D arrays of (b) refractive (pitch: $\unit{125}{\micro\meter}$) and (c) diffractive (pitch: $\unit{125}{\micro\meter}$) spherical microlenses; (d) 2D spot pattern in the re-imaged and demagnified focal plane and (e) fluorescence image of $^{85}$Rb atoms trapped in a section of the resulting array of micro-potentials representing 100 individually addressable qubits (pitch: $\unit{13}{\micro\meter}$, waist: $\unit{1.6}{\micro\meter}$, image is averaged 149 times).}
\label{fig:microoptics}
\end{figure*}%
%
The ability to synchronously investigate multi-component quantum systems decoupled from the environment in multi-site architectures  is fostering some of the most active research in quantum physics and quantum information processing \cite{2000:QCQ:544199}. Among the many currently pursued ap\-proa\-ches which range from solid state physics to quantum optics \cite{Bouwmeester:2001,Beth:2005:QIP:1076259,Everitt:2005:EAQ:1205130,Schleich:2007:EQI:1526289}, the ones in atomic physics seem to be particularly suited for advancing the field at this stage. This is due to the remarkable experimentally achieved control of single and multiple qubit systems, of qubit interactions, and the detailed understanding and control of the relevant coherent and incoherent processes, including excellent decoupling from the environment. Concentrating on work with neutral atoms, recently there has been a series of important advances, such as the near-deterministic preparation of single atomic qubits \cite{2001Natur.411.1024S,2011NatPh...6..951G}, the coherent transport of atomic quantum states \cite{2007NatPh...3..696B,PhysRevLett.91.213002,PhysRevLett.105.170502}, the manipulation of selected individual spins \cite{PhysRevLett.93.150501,PhysRevA.81.060308,2011Natur.471.319W}, and the implementation of two-qubit gates \cite{2003NatMandel,2007Natur.448..452A,PhysRevLett.104.010502,PhysRevLett.104.010503}.\\
Each of these experimental achievements represents an important step towards a successful physical implementation of quantum information processing \cite{2000ForPh..48..771D} by means of atom optics. Of great importance for future progress is consequentially the implementation of an architecture which incorporates all of the above achievements while at the same time providing scalability, reconfigurability, stability, and a modern technological basis as met for example by the newly emerging field of miniaturized and integrated atom optics \cite{2007BirklLasphotrev}.\\
This can be obtained by using different types of micro-fabricated configurations: the trapping and guiding of neutral atoms in micro-fabricated charged and current carrying structures have been pursued by a number of groups in recent years \cite{0022-3727-32-18-201,2002AdAMP..48..263F,RevModPhys.79.235,Reichel:2011:AtomChips}.
An alternative approach to generate miniaturized and integrated atom optical systems has been introduced by our group: we proposed \cite{2001OptComm,Buchkremer:2001} and demonstrated \cite{PhysRevLett.105.170502,PhysRevA.81.060308,PhysRevLett.89.097903,2007ApPhB..86..377L} the application of micro-fabricated optical elements (Fig. \ref{fig:microoptics}) for the manipulation of atomic qubits with laser light.
Using these elements for quantum information processing takes advantage of the vast industrial and research interest in the field of applied optics directed towards the development of micro-optical systems \cite{Jahns:2004:MTA:994002,2011:FMO:Zappe,herzig1997micro} and establishes a novel technological basis for research in quantum physics.\\
A special attraction of this approach lies in the fact that many of the currently used techniques in atom manipulation are based on the interaction of atoms with light. Thus, the use of micro-fabricated optical elements is in many ways the canonical extension of the conventional optical methods into the micro-regime, so that much of the knowledge and experience that has been acquired in macroscopic atom optics can be applied to this new regime in a very straightforward way. Moreover, the flexibility of the manufacturing process allows one to realize complex optical systems like computer-generated diffractive optical elements which can create light fields not achievable with standard components, and almost arbitrary spatial intensity distributions become possible. In addition, miniaturization enables one to scale from a single conventional element to multiple realizations, simply by utilizing parallelized lithographic fabrication techniques adapted from semiconductor processing.
The use of these manufacturing techniques allows the optical engineer to fabricate structures with dimensions in the micrometer range and submicrometer features, such as the diffractive lenses of Fig. \ref{fig:microoptics} (c). Up to $10^4$ microoptical elements can be produced on an area of $\unit{1}{\milli\meter\squared}$ while maintaining diffraction limited performance with numerical apertures (NA) large enough to define light patterns with structure sizes in the single micrometer regime.\\
Fig. \ref{fig:microoptics} shows typical examples of the micro-fabricated optical elements we use, the generated light fields, and fluorescence images of atoms trapped in the resulting potential geometries. One-dimensional arrays of cylindrical microlenses (Fig. \ref{fig:microoptics} (a)) allow us to realize atomic waveguides and arrays of interferometer-type guiding structures \cite{PhysRevLett.89.220402,PhysRevLett.92.163201}. Two-dimensional arrays of up to $300\times 300$ refractive (Fig. \ref{fig:microoptics} (b)) and diffractive (Fig. \ref{fig:microoptics} (c)) spherical microlenses are used to create 2D arrays of laser foci (Fig. \ref{fig:microoptics} (d)) which serve as 2D dipole trap arrays for neutral atoms with well over $100$ occupied sites (Fig. \ref{fig:microoptics} (e)) and typical site-to-site separations ranging from a few to about $100$ micron. 

\section{Scalable architecture for a neutral atom quantum processor}
\label{sec:architecture}

For a functional quantum processor, a sequential but partially also parallelized algorithm has to be implemented in a suitable geometry for performing the designated computational task: (a) qubits have to be prepared and initialized, (b) one- and two-qubit quantum operations have to be applied according to the quantum algorithm to be processed, and (c) high-fidelity readout of the final quantum state has to be achieved. An essential ingredient is a suitable architecture for the reliable storage and manipulation of qubits, thus presenting the hardware of the quantum processor.\\
Significant progress towards the implementation of this hardware has been achieved in systems relying on the optical storage of neutral-atom qubits, such as optical lattices \cite{PhysRevLett.93.150501,2011Natur.471.319W,2003NatMandel,2007Natur.448..452A,2009NatPh...5..575L,2007NatPh...3..556N,2009:Greiner:Microscope} or small configurations of individually focused laser beams \cite{2001Natur.411.1024S,PhysRevLett.104.010502,PhysRevLett.104.010503,PhysRevLett.96.063001}.
In our work, we have developed a quantum processor hardware based on the combination of optical methods for storage and control of neutral-atom qubits and the above introduced micro- and nano-fabricated optical systems, simultaneously targeting the important issues of single-site addressability and scalability. As a guideline for our work, we followed the generally acknowledged requirements for the physical implementation of quantum computing, as for example listed in reference \cite{2000ForPh..48..771D}. In specific, we have developed a scalable architecture for quantum information processing based on 2D quantum state registers built from 2D arrays of optical micro-potentials as shown in Figs. \ref{fig:microoptics} (d) and (e) and Fig. \ref{fig:architecture}:
%
\begin{figure*}
 \begin{center}
   \includegraphics[width=0.5\textwidth]{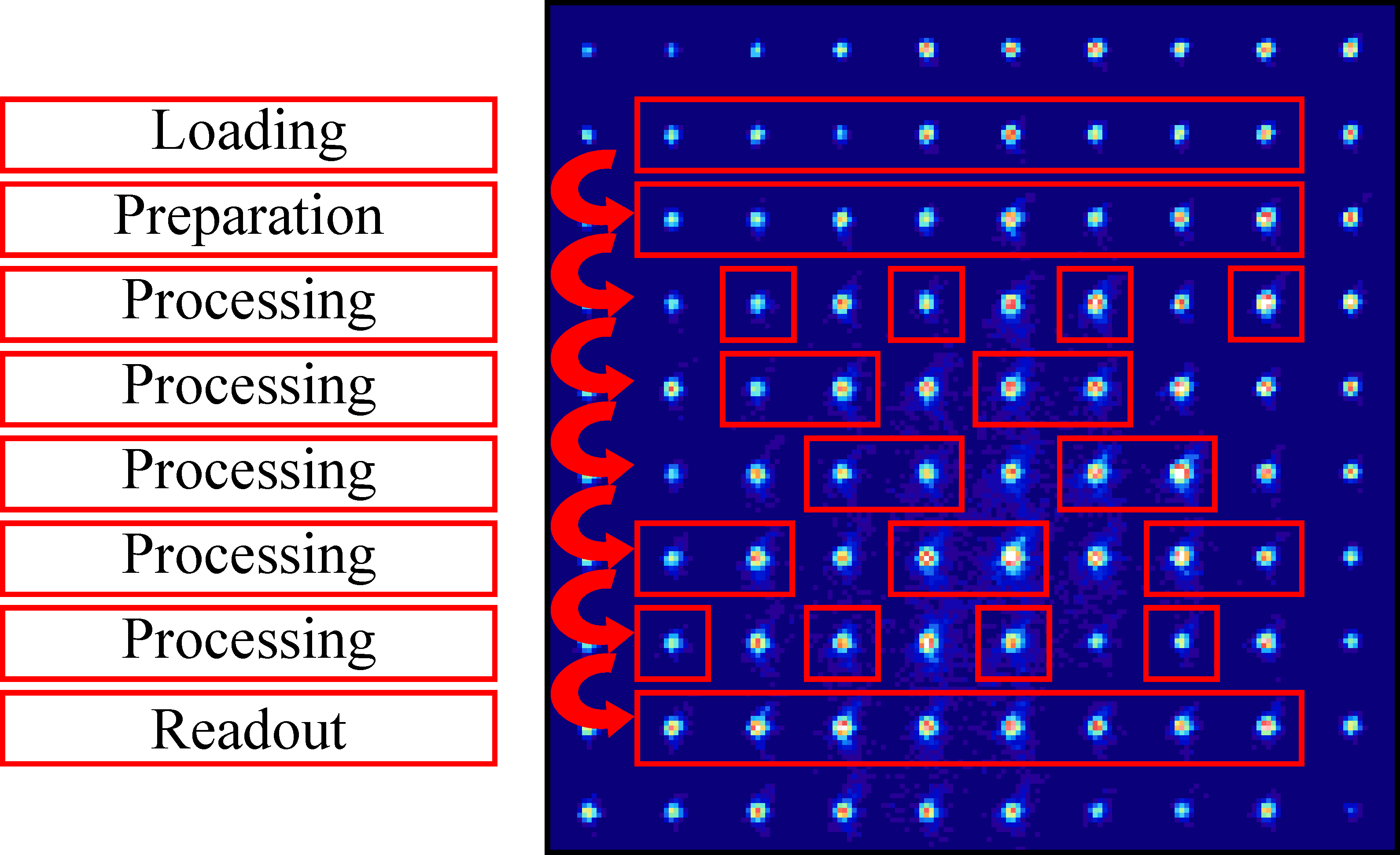}
 \end{center}
\caption{Quantum processor architecture based on a quantum shift register realized by a 2D array of optical micro-potentials. The array of single-addressable register sites serves as processor hardware with spatially separated loading, preparation, processing, and readout sections (left). Shift operations (arrows) are used to transport qubits through the architecture and selectively addressable operational units (red frames at right) are used to perform the individual steps of a quantum algorithm.}
\label{fig:architecture}
\end{figure*}%
%
\begin{itemize}
\item[$\bullet$] 2D configurations of laser beams focused by 2D arrays of microlenses serve as registers of optical potentials for the storage of small samples or single neutral atoms ($^{85}$Rb in our case), thus laying the foundation of a quantum register based processor architecture, where quantum information can be inscribed in the internal or external atomic states. Possible implementations range from small size registers where the sequential algorithm is applied in a temporal sequence to a localized set of qubits to large-scale registers with spatially separated functional subsections (see Fig. \ref{fig:architecture}) where atomic qubits or even atomic quantum bytes are transported during the algorithm resembling a standard shift register operation (see Section \ref{sec:storage}).
\item[$\bullet$] The reliable operation of the quantum processor requires the precise initialization and readout of each qubit together with targeted single-qubit and two-qubit gate operation. This requires the ability to perform incoherent and coherent operations in a global but also in a  site-specific fashion which is one of the inherent advantages of our architecture (see Section \ref{sec:prepmanread}).
\item[$\bullet$] A high degree of flexibility in the architecture is neccessary to implement different algorithms and to perform the sequential operations within an algorithm efficiently. The combination of microlens arrays with reconfigurable spatial light modulators allows us to implement adaptable trap configurations and reconfigurable schemes for qubit manipulation (see Section \ref{sec:slm}).\pagebreak
\item[$\bullet$] 
The implementation of a quantum shift register operation realizes the data bus in our architecture. It connects adjacent operational units, e.g. loading, preparation, processing, and readout sections (Fig. \ref{fig:architecture}). During operation, qubits are shifted in parallel through the quantum processor, allowing for massively parallelized quantum information processing. The preservation of coherence during the shift operation becomes an essential factor in evaluating this approach (see Section \ref{sec:transport}).
\item[$\bullet$] In addition to discussing the state-of-the-art of our architecture, we show that there is a well-defined and straightforward path for implementing the last remaining - but nevertheless crucial - element still missing for a functional quantum processor: two-qubit-gate operations. Several potential schemes have been proposed which can be implemented in our architecture (see Section \ref{sec:gates}).
\end{itemize}
%
\begin{figure*}
 \begin{center}
   \includegraphics[width=0.75\textwidth]{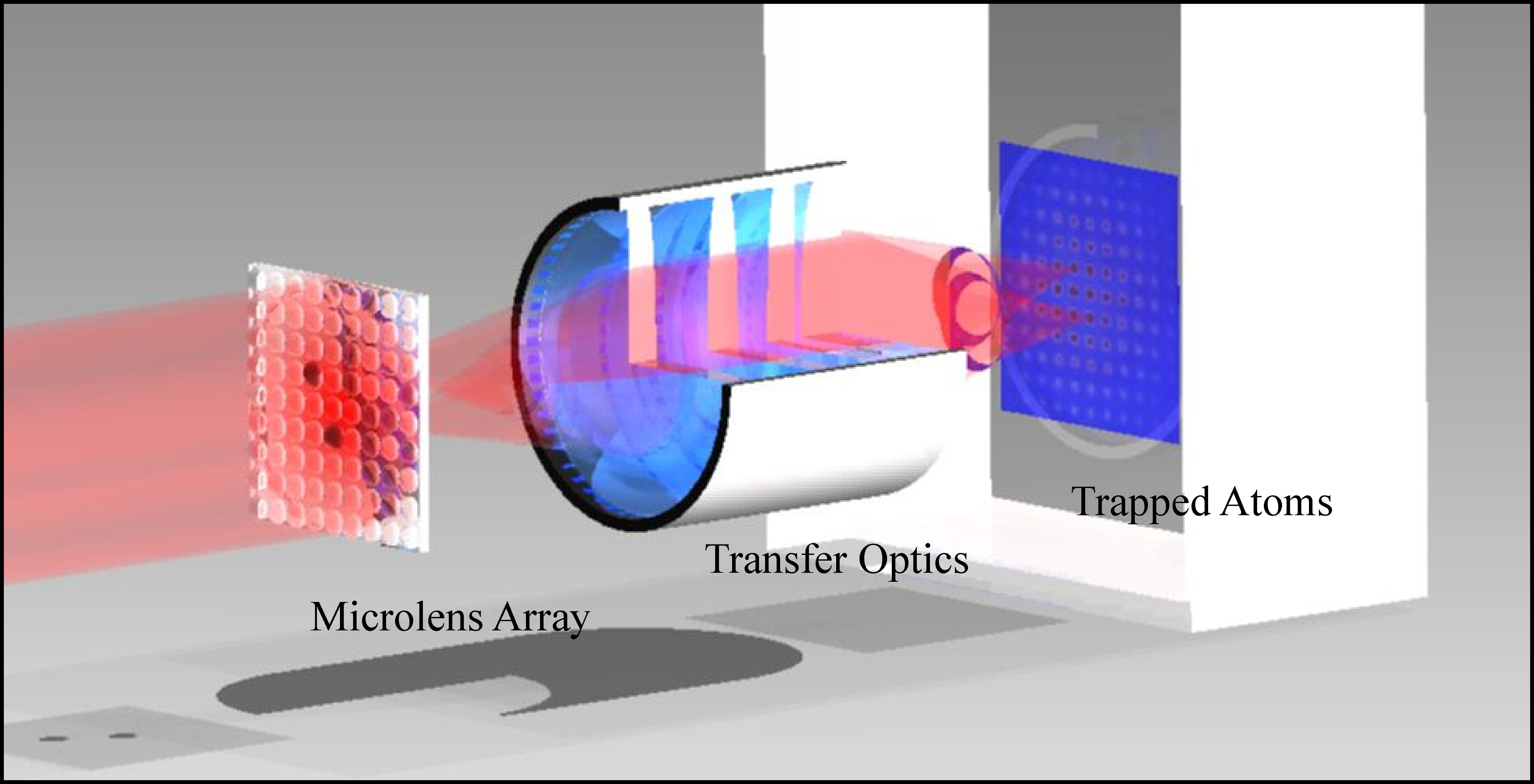}
 \end{center}
\caption{Schematic of the experimental setup of a scalable quantum processor hardware for neutral atoms: the trapping laser globally illuminates a microlens array which produces a 2D array of focal spots. After re-imaging the focal plane into a vacuum chamber, rubidium atoms are loaded into the resulting 2D dipole trap array, thus creating a 2D neutral-atom quantum register.}
\label{fig:setup}
\end{figure*}
%
In the following sections, we discuss in detail how our architecture can meet the above listed requirements. The experimental setup of Fig. \ref{fig:setup} shows the central elements for the quantum processor hardware. The key element is a 2D microlens array which, globally illuminated with appropriate laser light, produces a 2D array of laser spots in the focal plane. The focal plane is re-imaged into a vacuum chamber by a demagnifying imaging system, thus producing a 2D register of optical traps for neutral atoms with typical trap separations ranging from single to about $100$ microns. Fully exploiting our maximum available NA of 0.29, a waist below \unit{1.3}{\mu m} could be reached. Each optical trap can hold an ensemble of up to 100 atoms or in the limiting case an individual atom, thus a 2D register of atomic qubits with excellent scaling properties is created. Already in the present realization, lens arrays with ten thousands of individual lenses are available, a number which is by far not at the limit of the available technology of micro-optics fabrication.

\section{Quantum state storage in optical micro-potentials}
\label{sec:storage}

\begin{figure*}
 \begin{center}
   \includegraphics[width=0.75\textwidth]{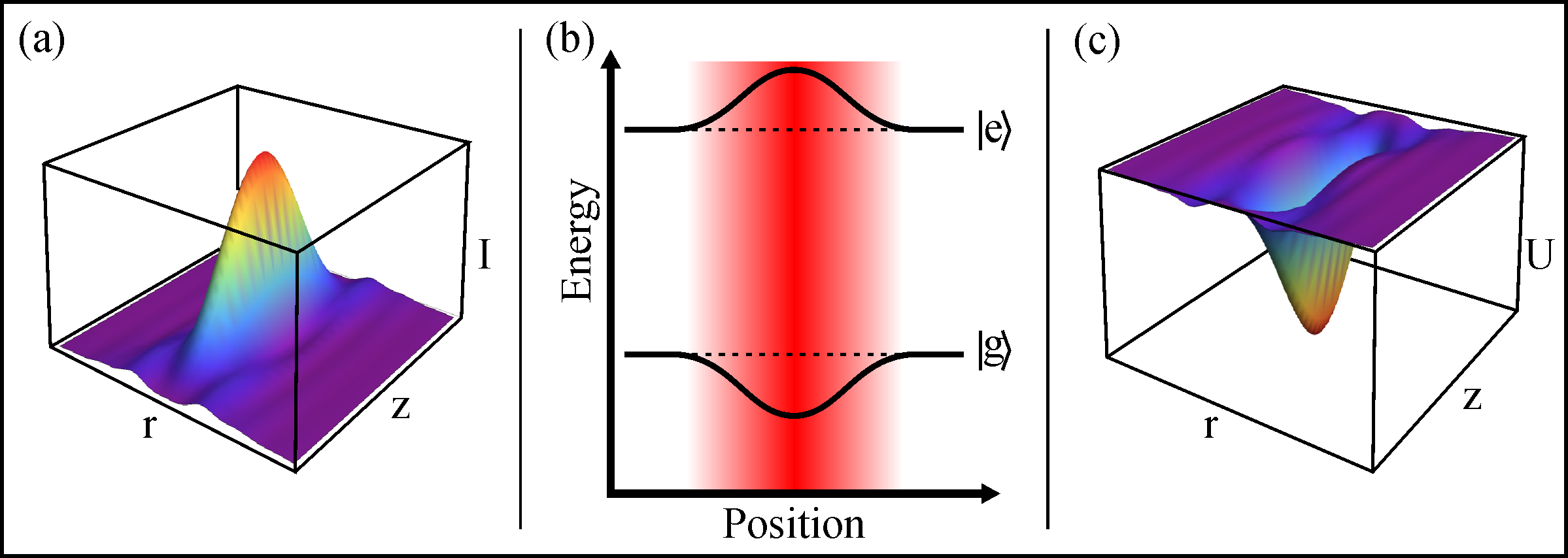}
 \end{center}
\caption{Principle of optical trapping of neutral atoms: (a) inhomogeneous intensity distribution of a focused laser beam; (b) energy shifts of the ground $\ket{g}$ and excited $\ket{e}$ state of an atomic two-level system induced by a red-detuned light field; (c) three-dimensional potential minimum for groundstate atoms in the focused laser beam of (a). See text for details and typical parameters.}
\label{fig:dipole}
\end{figure*}

Atomic quantum systems offer the important advantage that they can be localized and cooled at predefined sites as well as decoupled from their environment to a high degree. For neutral atoms, as an alternative to using the energy shift in magnetic fields \cite{0022-3727-32-18-201,2002AdAMP..48..263F,RevModPhys.79.235,Reichel:2011:AtomChips}, this can be achieved by using the energy shift in inhomogeneous optical fields \cite{Dalibard:85,Metcalf1999,grimm-2000-42}. Here, the short range character of the trapping force additionally facilitates the decoupling from the environment.\\
Optical trapping in dipole traps relies on the modification of the atomic energies by far-detuned laser light which is commonly described through the interaction part of the Hamiltonian 
\begin{equation}
H=H_{atom}+H_{light}+H_{int}.
\end{equation}
In most of the relevant cases, the atomic Hamiltonian $H_{atom}$ can be restricted to atomic two-level systems and the interaction Hamiltonian $H_{int}$ to atom-light coupling in dipole approximation. The resulting energy shift $\Delta E$ leads to the position dependent dipole potential $U\left(\mathbf{r}\right)$ for atoms in the ground state and a corresponding photon scattering rate $\Gamma_{SC}\left(\mathbf{r}\right)$ of
\begin{equation}
U\left(\mathbf{r}\right)=\frac{3\pi c^2}{2\omega_0^3}\frac{\Gamma}{\Delta}I\left(\mathbf{r}\right)\quad;\quad\Gamma_{SC}\left(\mathbf{r}\right)=\frac{3\pi c^2}{2\hbar\omega_0^3}\left(\frac{\Gamma}{\Delta}\right)^2 I\left(\mathbf{r}\right)
\end{equation}
with the rotating wave approximation applied. Here $I\left(\mathbf{r}\right)$ is the position-dependent laser intensity of the focused laser beam with waist $w_0$ ($1\slash e^2$-radius) and $\Delta=\omega_L-\omega_{eg}$ is the detuning of the laser field with respect to the resonance frequency of the two-level system spanned by $\ket{g}$ and $\ket{e}$ (Fig. \ref{fig:dipole} (b)) having a natural linewidth $\Gamma$. 
%
%
The above equations exhibit the essential features in optical dipole trapping: The magnitude of the energy shift depends linearly on the trapping laser intensity $I\left(\mathbf{r}\right)$ at the position of the atom and its sign is given by sign of the laser detuning $\Delta$. Therefore, the inhomogeneous intensity profile of a focused Gaussian laser beam (Fig. \ref{fig:dipole} (a)) creates a reduction of the atom's groundstate energy (Fig. \ref{fig:dipole} (b)) and thus an attractive trapping potential with depth $U_0$ in three dimensions for red detuning $\Delta<0$ (Fig. \ref{fig:dipole} (c)). Furthermore, unwanted exitation to the excited state and resulting spontaneous scattering can be kept low for large detuning, since the dipole potential scales with $1\slash\Delta$, whereas the scattering rate scales with $1\slash\Delta^2$.\\
Important characteristics of these traps are potential depths of up to several $\unit{\milli\kelvin}{}\times\unit{k_B}{}$, which are about two orders of magnitude larger than the thermal energies achievable with standard laser cooling techniques \cite{Metcalf1999} and vibrational frequencies in the range of $\unit{10}{\kilo\hertz}$ to $\unit{100}{\kilo\hertz}$ or even beyond for tight focusing. By advanced laser cooling, e.g. Raman sideband cooling \cite{PhysRevLett.69.1741,PhysRevLett.80.4149}, or by making use of the phase transition to a Bose-Einstein condensate (BEC) \cite{RevModPhys.74.875,RevModPhys.74.1131}, the vibrational ground state of the trapping potentials can be populated with high probability. Corresponding spreads of the ground state wave functions are on the order of $\unit{10}{\nano\meter}$ for single atoms.\\
A typical set of parameters for the experiments with $^{85}$Rb atoms presented in the following sections of this work are a trapping laser wavelength of $\unit{815}{\nano\meter}$ (Ti:Sapphire laser) which corresponds to an effective detuning of about $\Delta=\unit{2\times 10^6}{\Gamma}$ with respect to the rubidium D-Lines at $\unit{780}{\nano\meter}$ and $\unit{795}{\nano\meter}$ and an optical power of $\unit{2}{\milli\watt}$ in the central focal spot of the microlens register. The implemented trap array has a pitch of $a=\unit{55}{\micro\meter}$, a waist of $w_0=\unit{3.7}{\micro\meter}$ and corresponding Rayleigh range of $z_R=\unit{52.8}{\micro\meter}$ which yields a trap depth of $U_0=\unit{k_B\times 0.1}{\milli\kelvin}$. The scattering rate evaluates to $\Gamma_{SC}=\unit{6}{s^{-1}}$. In addition, the coherence limiting state-changing part of photon scattering is suppressed by quantum interference effects \cite{1994OptL...19..207C} to a value of $\unit{0.5}{s^{-1}}$ which already in this configuration gives a limit to the coherence time of $2$ seconds. With the available laser power of $\unit{1}{\watt}$, several 100 sites of the processor architecture are accessible, but we limit the number of investigated qubits to about 100 in our current work (Fig. \ref{fig:microoptics} (e)). Changing to a trapping laser with even larger detuning, e.g. using light at $\unit{1064}{\nano\meter}$ (Nd:YAG-laser), a power of $\unit{14}{\milli\watt}$ in the central trap leads to the same trap depth and an absolute scattering rate of about $\unit{0.3}{s^{-1}}$. Here, unwanted state-changing scattering is suppressed to $\unit{3\times 10^{-4}}{s^{-1}}$. Again, typically available laser powers of $\unit{10}{W}$ lead to architectures with several 100 register sites, now having coherence times in the range of minutes.\\
Neutral atoms stored in this register represent intrinsically identical quantum systems which are decoupled from their environment to a high degree \cite{PhysRevLett.89.097903,2007ApPhB..86..377L}. In addition, there is a wide range of options for encoding quantum information in neutral atoms in this architecture: quasi spin-$1\slash 2$ systems can be generated in the external degrees of freedom \cite{PhysRevLett.90.147901,PhysRevA.66.042317,PhysRevA.70.023606,Eckert2006264}, e.g. in the vibrational modes of the trapping potential, as well in the internal degrees of freedom represented by two states of the hyperfine manifold of the electronic ground state of the trapped atoms, as shown in \cite{2007ApPhB..86..377L}.

\section{Initialization, readout, and 1-qubit-rotation}
\label{sec:prepmanread}

Alkali atoms - especially rubidium and caesium - have become the preferred atomic species for research in quantum information processing with neutral atoms: alkali atoms can be efficiently controlled by laser light in the external degrees of freedom as described in the previous section, but also in their internal states which is essential for quantum state preparation, manipulation and readout.\\
Optical pumping \cite{RevModPhys.44.169}, which is based on applying resonant laser light of adequate polarization, allows one to prepare atoms in desired internal states, e.g. the "clock states" ($\ket{F=2,m_F=0}$ and $\ket{F=3,m_F=0}$ of $^{85}$Rb) of the ground state hyperfine manifold of rubidium. The resulting true two-level quasi spin-$1\slash 2$ system is an excellent qubit basis. The pumping light can be applied globally or site-selectively in our architecture for efficient initialization of variable qubit configurations. Selective readout of the qubit state can be achieved by utilizing fluorescence imaging, which can be done spatially resolved with a CCD-camera, and therefore simultaneously for all sites of the qubit register (Fig. \ref{fig:microoptics} (e)) \cite{PhysRevLett.89.097903}.\\
%
\begin{figure*}
 \begin{center}
   \includegraphics[width=0.75\textwidth]{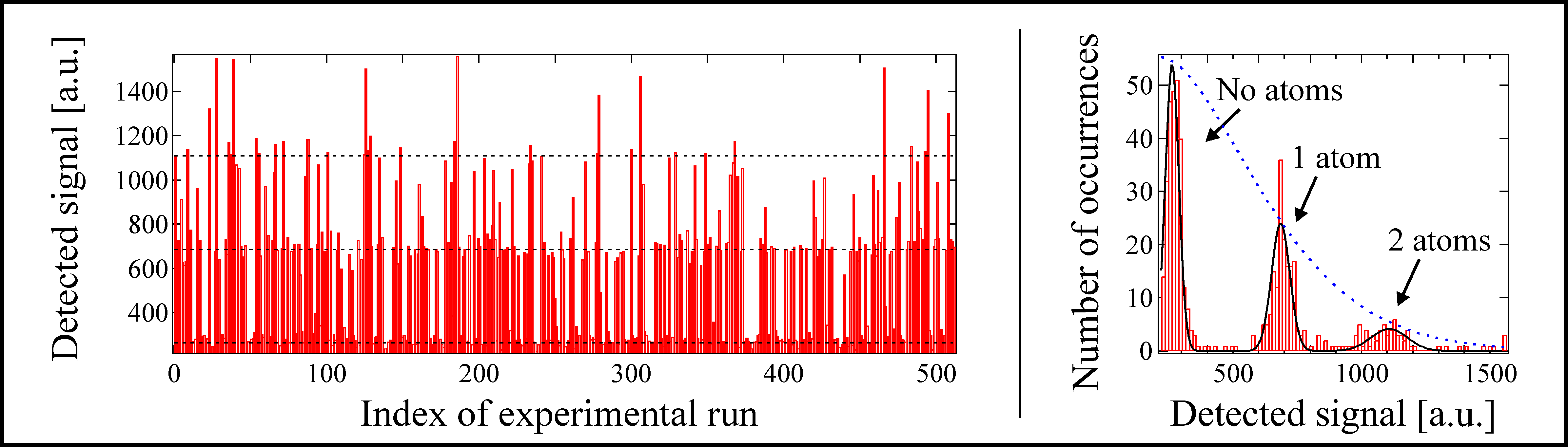}
 \end{center}
\caption{Fluorescence signal obtained from atoms at an individual register site (left) and corresponding histogram of the detected signal (right). Steps in the detected signal correspond to background light ($\approx 300$ a.u.), single-atom events ($\approx 700$ a.u.) and two-atom events ($\approx 1100$ a.u.).}
\label{fig:SA}
\end{figure*}%
%
In fluorescence imaging, the detected signal level corresponds to the atom number at each site since every atom contributes a comparable amount of photons to the signal. With low-noise detection schemes another important requirement for the successful implementation of quantum information processing can be fulfilled: the number of atoms at each register site can be determined precisely, especially at the single atom level. The characteristics of number resolved atom detection in our architecture are illustrated in Fig. \ref{fig:SA}. The left side displays the fluorescence signal obtained for one selected trap out of a 2D trap register for about $500$ consecutive experimental runs. The signal clearly exhibits reoccuring levels in signal amplitude, such that the levels for background light scattering (i.e. no atom), single-atom events, and two-atom events can be clearly discriminated. This becomes even more obvious in a histogram analysis of the experimental data (Fig. \ref{fig:SA} (right)) which exhibits distinct peaks for 0, 1, and 2 atoms. For a statistical loading process, as implemented in the experimental situation presented in Fig. \ref{fig:SA}, a Poissonian probability distribution for the atom number distribution is observed and the maximum probability for single-atom events is limited to $\unit{37}{\%}$, while two-atom events are present as well. More advanced loading schemes have been implemented in single dipole trap experiments \cite{2001Natur.411.1024S,2011NatPh...6..951G}, including the possibility of eliminating two-atom events and increasing the single-atom loading efficiency by utilizing light assisted collisions \cite{RevModPhys.71.1}. Single-atom probabilities of $\unit{50}{\%}$ in the regime of collisional blockade \cite{2001Natur.411.1024S} and up to $\unit{83}{\%}$ for an optimized process starting from an ensemble of atoms \cite{2011NatPh...6..951G} with no or almost no two-atom events have been achieved. Implementing these techniques also in our architecture should lead to a collective near deterministic preparation of single atoms at all sites of our qubit register.\\ 
%
\begin{figure*}
 \begin{center}
   \includegraphics[width=0.75\textwidth]{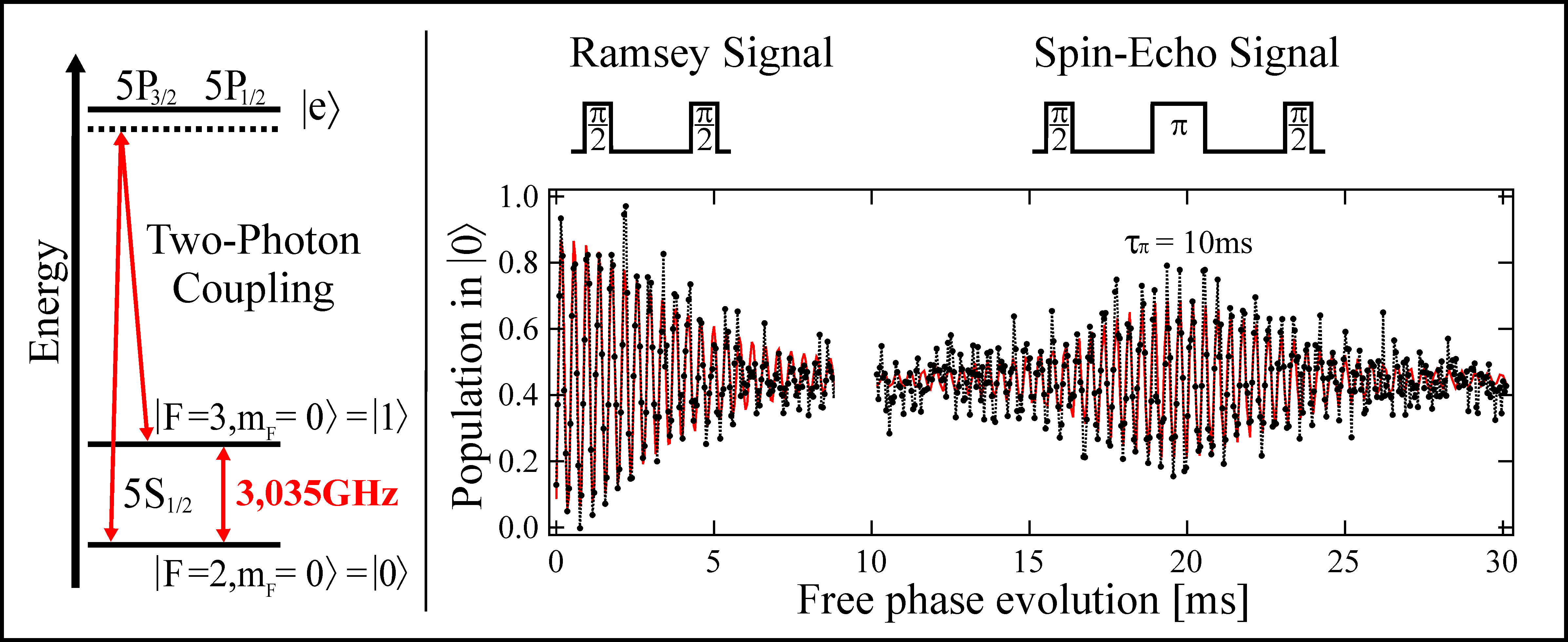}
 \end{center}
\caption{(left) Relevant part of the internal atomic structure of $^{85}$Rb and coherent optical coupling of the qubit basis states via two-photon transitions. (right) Ramsey and spin-echo dynamics of the qubit rotation. The observed decay of the signal amplitude corresponds to inhomogeneous (Ramsey signal) and homogeneous (spin-echo signal) dephasing.}
\label{fig:spectroscopy}
\end{figure*}%
%
For the sake of improved signal-to-noise ratio, all experiments presented in the following sections have been performed with small atom ensembles with atom numbers per site ranging from 10 to 100. The quantum state of the investigated atomic qubits, given by the superposition of the basis states $\ket{0}=\ket{F=2,m_F=0}$ and $\ket{1}=\ket{F=3,m_F=0}$ of $^{85}$Rb is accessible to full coherent control by microwave radiation or optically with single-site addressability through two-photon coupling via a virtually excited intermediate state (Fig. \ref{fig:spectroscopy} (left)). Achievable coupling strengths correspond to Rabi frequencies of several $\unit{10^6}{s^{-1}}$, thus allowing for spin rotations of $\pi$ on a microsecond timescale.\\
Fig. \ref{fig:spectroscopy} (right) displays a typical example for the coherent control of the quantum state dynamics in a Ramsey and spin-echo configuration \cite{2007ApPhB..86..377L}. In this experiment, the above described methods for quantum state preparation and control are utilized to analyze the time evolution of qubits initially prepared in a superposition of states $\ket{0}$ and $\ket{1}$. The signal amplitude decay in the Ramsey signal is due to reversible inhomogeneous dephasing which is eliminated in the spin-echo configuration. We label the time constant for homogeneous dephasing extracted from the latter case as decoherence time. In our architecture, decoherence times on the order of $\unit{100}{\milli\second}$ have been observed, already allowing for hundreds or thousands of coherent control pulses to be applied during qubit coherence \cite{PhysRevLett.105.170502}.

\section{Reconfigurable single-site addressable qubit register}
\label{sec:slm}

Architectures based on 2D arrays of tightly focused laser beams with typical separations in the micro\-meter regime for qubit storage inherently provide the ability to address the individual qubit sites since one can use the optics generating the trap array at the same time for addressing purposes.
Based on the scalable architecture presented above, we have introduced and experimentally implemented a novel approach for complementing the ability to perform quantum operations in parallel with an additional versatility by achieving reconfigurable, site-selective initialization and operation in freely selectable subsets of sites.
We combine 2D arrays of microlenses with per-pixel addressable spatial light modulators (SLM). This results in reconfigurable, per-site addressable 2D configurations of diffraction-limited laser foci in the focal plane of the microlens array which - as before - are re-imaged into the vacuum system \cite{PhysRevA.81.060308}. Central to our approach is the fact that we use the SLM only for the addressing of individual microlenses, but not as a holographic phase element for creating complex focal spot structures \cite{Bergamini:04,PhysRevA.73.031402}. This ensures high stability and a diffraction-limited performance, both given by the advantageous characteristics of the microlenses.\\
%
\begin{figure*}
\begin{center}
   \includegraphics[width=0.75\textwidth]{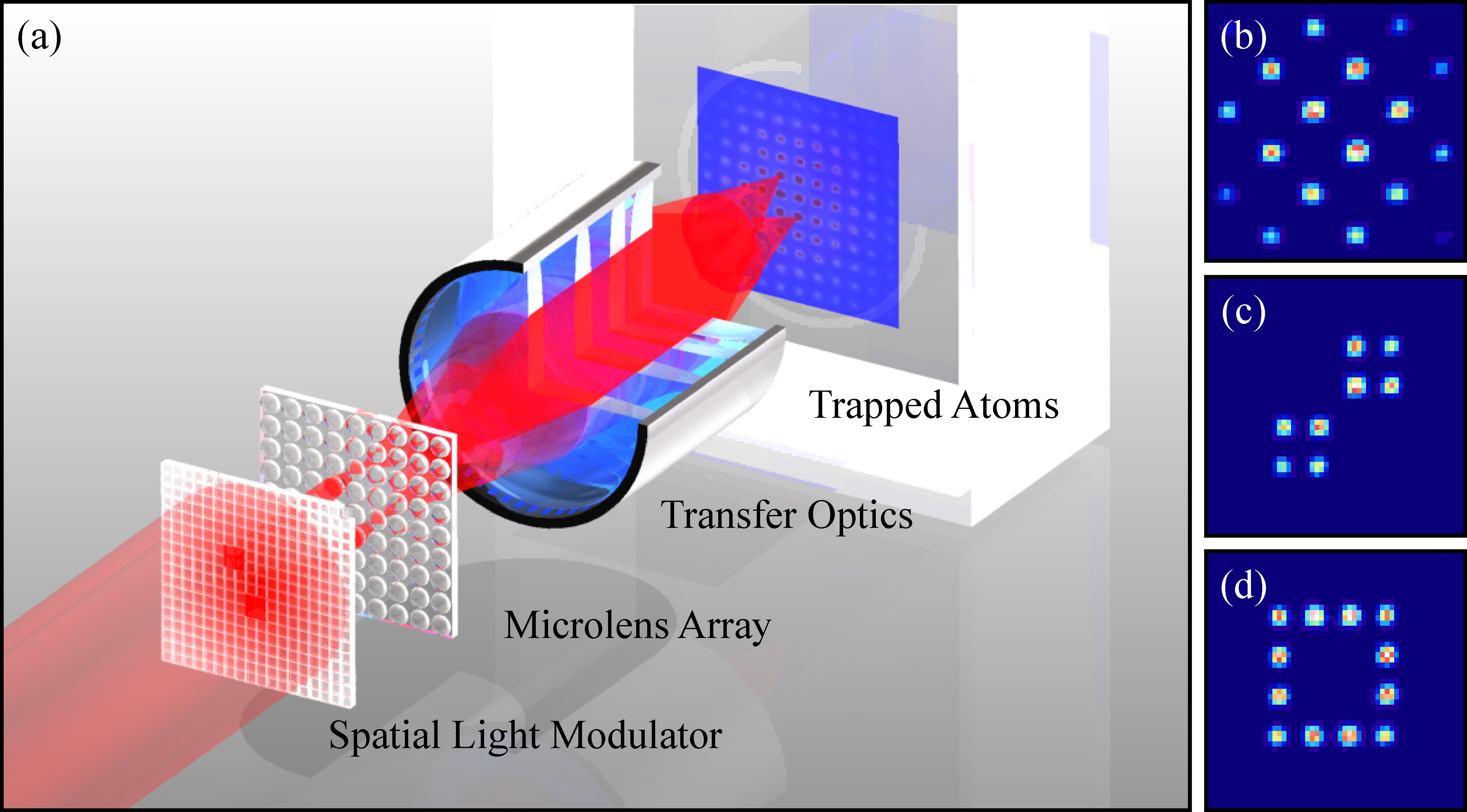}
 \end{center}
\caption{(a) Schematic experimental setup including a spatial light modulator (SLM) for addressing reconfigurable sections of the microlens array and (b-d) fluorescence images of atoms stored in different trap patterns (fundamental pitch: $\unit{55}{\micro\meter}$, images are averaged 20 times). See \cite{PhysRevA.81.060308} for details.}
\label{fig:addressing}
\end{figure*}%
%
A schematic view of the extended experimental setup is presented in Fig. \ref{fig:addressing}. Laser light for atom trapping or manipulation globally illuminates an SLM which is placed in front of the microlens array. The SLM is a 2D array of pixels, each acting as an individually tunable optical attenuator, which we use as a per-pixel intensity modulator. This allows one to separately control the light power impinging on each microlens by inscribing a reconfigurable pattern of transmitting or non-transmitting sections into the SLM. In the configuration used for the experiments presented below, an area of 80 pixels corresponds to one single microlens and the transmission of each pixel is subject to computer-control. This results in a range of the relative transmitted intensity between \unit{0.4}{\%} and \unit{100}{\%} corresponding to a contrast of 270:1. For the experiments presented here only static configurations of the SLM are used, but employing state-of-the-art fast updating devices, switching frequencies in the several kilohertz regime are achievable for liquid crystal or micro-mirror (DLP) based devices.\\
%
\begin{figure*}
 \begin{center}
   \includegraphics[width=0.75\textwidth]{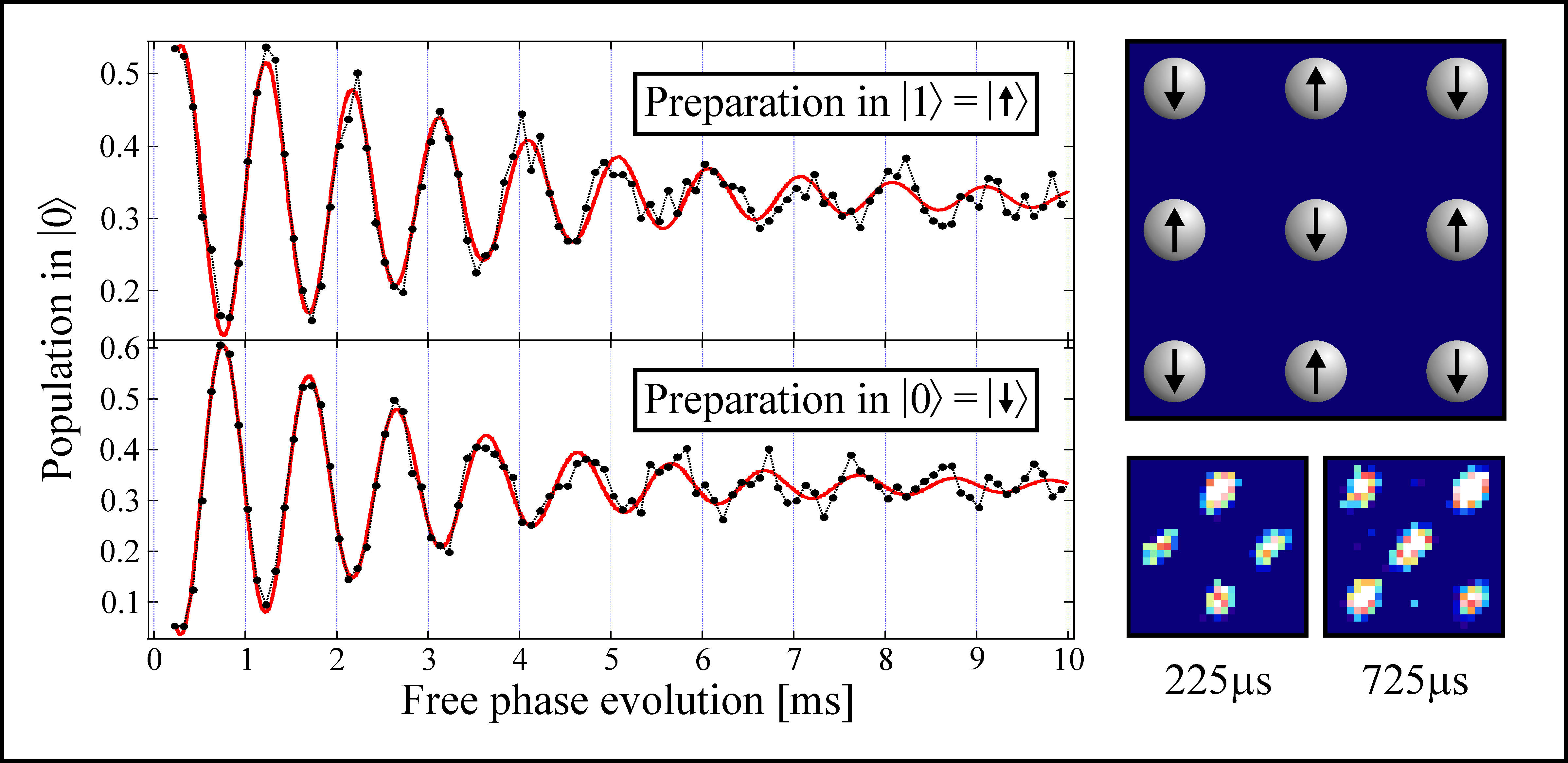}
 \end{center}
\caption{(left) Ramsey oscillations for atoms at two neighboring sites after initialization in opposite spin states according to a checkerboard pattern (right, top). (right bottom) Fluorescence images of atoms in state $\ket{0}$. Atoms at complementary sites are visible at different times due to the phase difference of $\pi$ between the two initially prepared opposite spin states. See \cite{PhysRevA.81.060308} for details.}
\label{fig:spinflip}
\end{figure*}%
%
We used this setup to produce versatile 2D configurations of atom traps \cite{PhysRevA.81.060308} as shown in Fig. \ref{fig:addressing}. Starting from the fundamental structure of the 2D trap register, created by globally illuminating the full microlens array (all pixels of the SLM turned to full transmission) we have demonstrated the ability to change the pitch and orientation of the qubit register by illuminating only every other microlens creating a 'superlattice' with definable structure (Fig. \ref{fig:addressing} (b)), to generate subsets of separated dipole trap arrays (Fig. \ref{fig:addressing} (c)), e.g. for the implementation of quantum error correction schemes or plaquette states in 2D lattice spin models \cite{lattice_spin_2006}, and to realize the structure of a ring lattice with periodic boundary conditions (Fig. \ref{fig:addressing} (d)) \cite{PhysRevLett.95.063201,PhysRevA.79.043419}.\\
In addition to creating flexible trap geometries, we also perform coherent manipulation of 2D sets of atomic quantum systems in parallel as well as site-selectively in a reconfigurable fashion. We use the combined system of SLM and microlens array in a very similar fashion as before but now for the control of the light inducing the two-photon coherent coupling. This provides fully flexible quantum state control of the qubits stored in the register by inscribing freely configurable phase shifts at definable sites \cite{PhysRevA.81.060308}.\\
This site-selective addressability allows one to prepare complex 2D spin configurations. As an example, we use the SLM to prepare a 2D configuration of alternatingly anti-parallel spins in neighboring trapping sites (Fig. \ref{fig:spinflip} (right, top)) by applying a $\pi$ phase shift in the pattern of Fig. \ref{fig:addressing} (b) to atoms initially in state \ket{0} at all sites. To demonstrate the coherence of this site-selective reversal of spins, a Ramsey experiment is performed in all traps simultaneously after the spin-flip operation. In Fig. \ref{fig:spinflip} (right, bottom) two fluorescence images showing atoms in state \ket{0} after different free evolution times are presented for nine traps and Ramsey oscillations in two neighboring traps are given in detail (Fig. \ref{fig:spinflip} (left)). All traps show Ramsey oscillations, but due to their different initial spin states, we observe the expected phase difference of $\pi$ in the Ramsey oscillations between qubits initially prepared in $\ket{0}$ and $\ket{1}$, respectively.

\section{Coherent transport of atomic quantum states}
\label{sec:transport}

\begin{figure*}
 \begin{center}
   \includegraphics[width=0.75\textwidth]{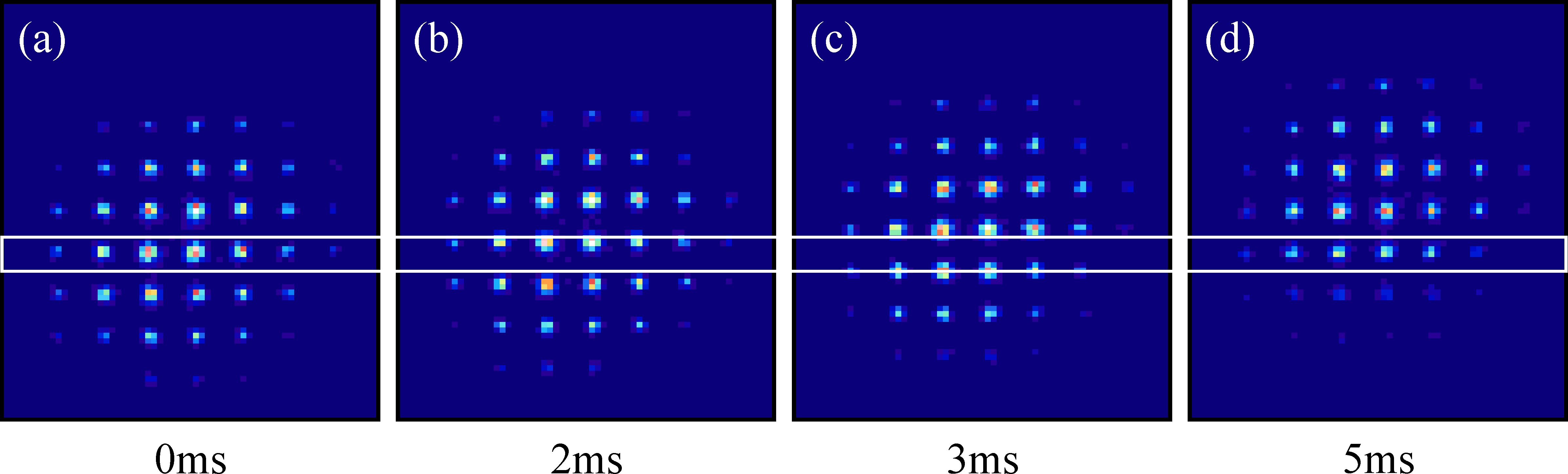}
 \end{center}
\caption{Fluorescence images of atoms transported over a distance of one full trap separation ($a=\unit{55}{\micro\meter}$) in a shift register: (a) initial position, (b-c) positions after $\unit{2}{\milli\second}$ and $\unit{3}{\milli\second}$, respectively, and (d) position after a shift over one full trap separation ($\unit{5}{\milli\second}$).}
\label{fig:transport}
\end{figure*}

\begin{figure*}
 \begin{center}
   \includegraphics[width=0.75\textwidth]{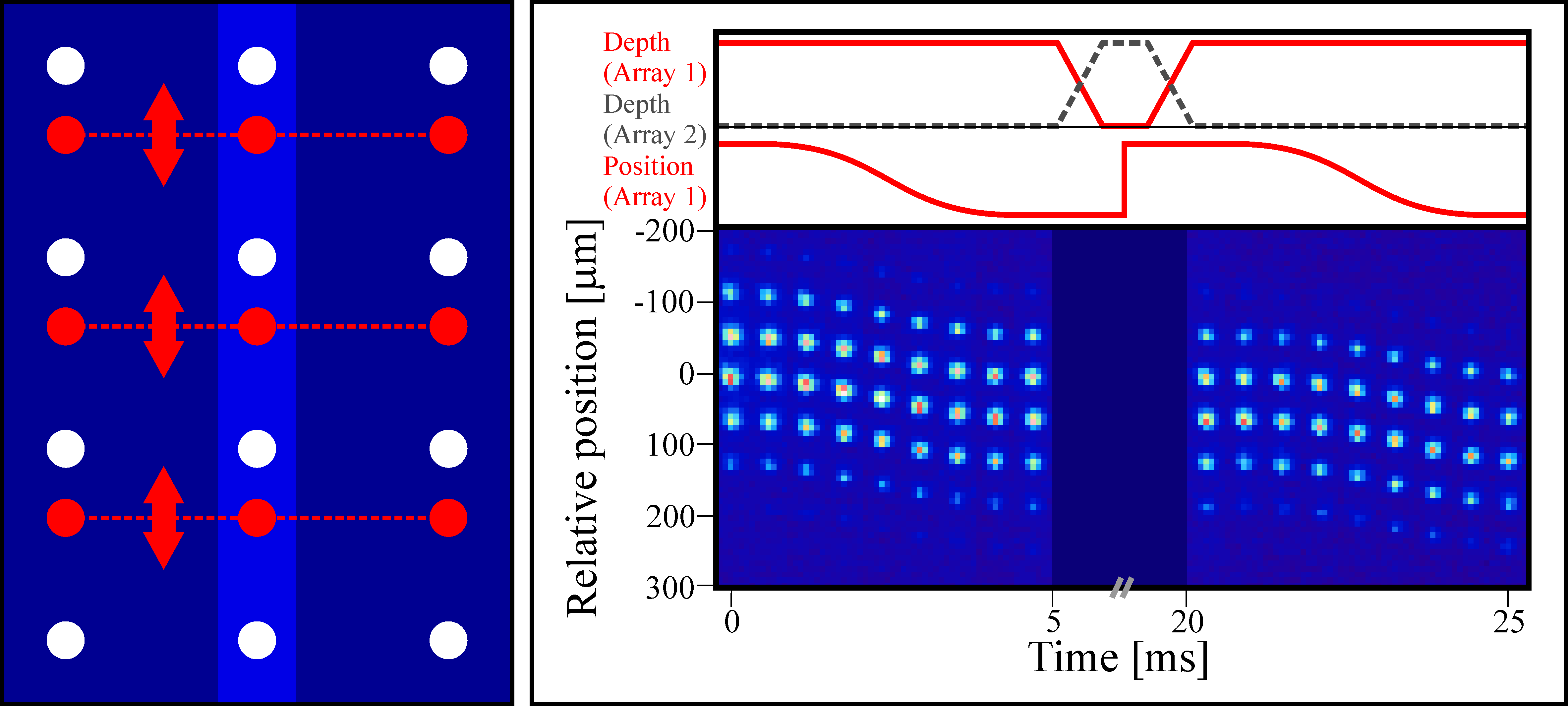}
 \end{center}
\caption{Scalable shift register. (left) Two superimposed dipole trap arrays (one moveable (red) and one static (white)) give control over trap separations and scalable atom transport. Fluorescence images of atoms (right bottom) in the central column (as indicated by the brighter column at left) of a 2D shift register  during two consecutive shift sequences together with the corresponding timing sequences for trap depths and position (right top). See \cite{PhysRevLett.105.170502} for details.}
\label{fig:ribbon}
\end{figure*}

\begin{figure*}
 \begin{center}
   \includegraphics[width=0.75\textwidth]{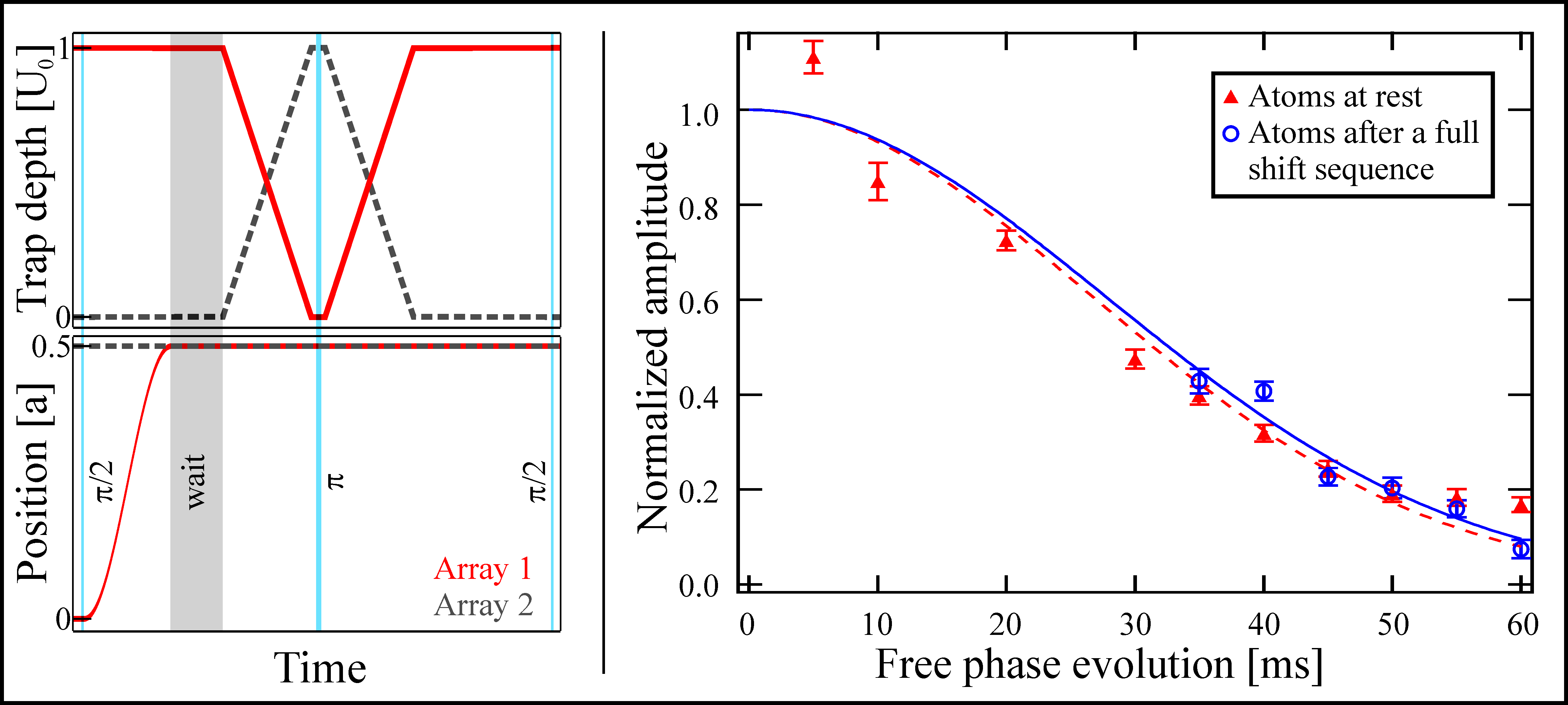}
 \end{center}
\caption{Preservation of coherence during the shift register operation. (left) Timing sequence for the spin-echo measurement applied for a determination of the homogeneous dephasing time constant. (right) The signal decay due to homogeneous dephasing for atoms at rest (red triangles) and for atoms participating in a full shift register sequence (blue circles) is almost identical. Thus the shift register operation does not introduce additional decoherence. See \cite{PhysRevLett.105.170502} for details.}
\label{fig:coherenceSHIFT}
\end{figure*}

Central to the functionality of our complex processor architecture (Fig. \ref{fig:architecture}) is the implementation of a scalable quantum shift register which serves as data bus and connects spatially separated loading and processing units. In the following, we present an all optical device which offers precise control of the position and transport of trapped neutral-atom qubits in registers of dipole potentials. Moreover, this quantum shift register can serve as a 2D quantum memory to archive and retrieve quantum information, and sequentially shuffle quantum information through complex architectures \cite{PhysRevLett.105.170502}.\\
The shift operation is based on consecutive loading, moving, and reloading of qubits stored in two independently controllable quantum registers. This configuration is obtained from two superimposed dipole trap arrays created either from two separated microlens arrays or by irradiating a single microlens array with two trapping laser beams under different incident angles. To move the traps, we vary the incident angle of one of the trapping laser beams by a scanning mirror, which causes the foci of the respective array to shift laterally within the focal plane. Atoms stored in the trap register are transported along with the laser foci and it is straightforward to shift the array of trapped atoms by a distance of the full trap separation of $a=\unit{55}{\micro\meter}$ as shown in Fig. \ref{fig:transport}.\\
%
%
For a scalable shift register \cite{PhysRevLett.105.170502}, we combine the moveable quantum register with a static one of identical parameters (Fig. \ref{fig:ribbon} (left)). Consecutive moving and reloading of atoms between the two registers allows for atom transport over macroscopic distances, where the number of achievable shift sequences is only limited by the size of the illuminated trap array (Fig. \ref{fig:transport}). The fluorescence images in Fig. \ref{fig:ribbon} (right) show two shift sequences in detail. Pictured is the central column of a 2D quantum register (as indicated by the differently colored central colunm in Fig. \ref{fig:ribbon} (left)) as a function of time together with the corresponding timing sequence for the depths and positions of the two dipole trap arrays. The shift operation is performed as follows: atoms are loaded into the moveable trap register (Array 1), shifted for a full trap separation, and transferred to the static register (Array 2) where they are stored while the moveable register is returned to its initial position. To complete a shift cycle, the atoms are loaded from the static register back to the moveable one, ready for a repetition of the shift sequence.\\
We do not observe any atom loss or heating when reloading between identical potential wells or during transport with durations in the single millisecond regime, which ensures the ability of high-fidelity transport of atoms over macroscopic distances for sufficiently large trapping arrays. Technical optimization is capable of pushing time constants in this process below the millisecond regime to the limit given by vibrational frequencies of the trapping potential.\\
An essential requirement for quantum information processing in this architecture is the preservation of coherence during transport, reloading, and the full shift sequence. We have performed a detailed investigation on the influence of the shift register operation on coherence in order to address that issue \cite{PhysRevLett.105.170502}. We embedded the shift register cycle in a spin-echo experiment, thereby analyzing its influence on the decoherence time. The corresponding experimental sequence is shown in Fig. \ref{fig:coherenceSHIFT} (left).
%
%
For a quantitative investigation, Fig. \ref{fig:coherenceSHIFT} (right) presents the signal contrast, i.e. the maximum amplitude $A_{\textrm{E}}(2t_{\pi})$ of the echo signal at time $2t_{\pi}$ normalized to the amplitude of the Ramsey signal $A_{\textrm{R}}(t=0)$ as a function of the free evolution time $2t_{\pi}$ for atoms at rest (red triangles) and atoms participating in the full shift register sequence (blue circles). The loss in signal contrast is almost identical in both cases. From a detailed analysis of external influences, we determine homogeneous dephasing due to irreversible variations of the atomic resonance frequency to be the dominant cause for loss of contrast. We identify heating caused by photon scattering from the trapping laser to be the most likely cause for this. Following the calculations given in \cite{PhysRevA.72.023406}, the signal contrast should be described by the Gaussian function \mbox{$C(2t_{\pi})=C(0)\exp(-(2t_{\pi})^{2}/T_{2}'^2)$} with time constant $T_{2}'$ for reduction of the initial contrast to its $e^{-1}$-value. The measurements in Fig. \ref{fig:coherenceSHIFT} can be well fitted to $C(2t_{\pi})$ which gives time constants for both cases (atoms at rest and atoms in the shift register) of about $\unit{40}{\milli\second}$. On average, the ratio of experientially determined coherence times evaluates to $T_{2,shift}'\slash T_{2,rest}'=0.98 (4)$. Thus, no additional dephasing or decoherence of internal-state superposition states occur for the full shift register cycle within the measurement uncertainty. This proves that the qubit transport (as necessary for most of the two-qubit gate operations proposed in Sec. \ref{sec:gates}) preserves coherence and that the fundamental shift sequence can be cascaded and thus scaled to complex and versatile 2D architectures allowing coherent quantum state storage and transport along complex and reconfigurable paths.
%
%
\section{Prospects for the implementation of two-qubit gates}
\label{sec:gates}

One of the essential requirements for the realization of a quantum processor is the capability of performing arbitrary one-qubit gates and at least one suitable two-qubit gate \cite{2000ForPh..48..771D}. Together they represent a universal set of quantum gates \cite{PhysRevA.52.3457}.
Our single-site addressable 2D quantum register inherently provides the framework for single-qubit operations as presented in Secs. \ref{sec:prepmanread} and \ref{sec:slm}.\\
The realization of two-qubit gates, in general, is still subject to active research in a variety of approaches towards the physical implementation of quantum computation and is inevitably linked to great demands on experimental control, since interaction becomes indispensable.
There are several promising proposals for architectures based on optically trapped neutral atoms in multi-site potential wells.\\
Originating from the rapid developments in the field of ultracold quantum gases, the idea of entangling atoms via controlled s-wave collisions was formulated \cite{PhysRevLett.82.1975}. The deterministic implementation of coherent collisional events relies on the precondition of preparing single atoms in well-defined vibrational states (e.g. the ground state) of the trapping potential which is characteristic for the transition to a Mott-Insulator state in a BEC. Hence, remarkable experimental results have been achieved in optical lattices, namely the production of entangled states \cite{2003NatMandel} and the realization of a two-qubit $\sqrt{SWAP}$ phase gate \cite{2007Natur.448..452A}. Both realizations employ time and state dependent potentials for the controlled overlap and separation of atomic wave functions. In this regard, the coherent transport of atomic quantum states (Sec. \ref{sec:transport}) with full control over site separation represents a fundamental step which has to be complemented with state-selectivity and cooling to the ground vibrational state in order to utilize cold collisions according to the above method in the current setup.\\
In addition, the shift register is the foundation for two-qubit-gate proposals based on the external degrees of freedom. One is based on the use of the two lowest vibrational levels as qubit basis where gate operations involve tunnelling controlled by adiabatic spatial approach and separation of traps together with cold collisional interaction \cite{PhysRevA.66.042317,PhysRevA.70.023606,Eckert2006264}, and a second one is based on quantum computing with spatially delocalized qubits \cite{PhysRevLett.90.147901}. In the latter case, the computational basis states are defined by the presence of a single atom in the ground state of one out of two trapping sites. There is a fascinating extension of this approach, also making use of the underlying concept of adiabatically connecting adjacent trapping sites: "Atomtronics" with holes \cite{PhysRevA.82.013604} allows for the construction of a coherent single hole diode and transistor in an array of dipole traps filled with two atoms and one hole. These elements could serve as the fundamental building blocks for atom-based implementations of processor schemes analogous to currently used electronic devices, but working on coherent principles and single-atom control.\\
Through recent experimental success \cite{PhysRevLett.104.010502,PhysRevLett.104.010503,PhysRevA.82.030306}, two-qubit gates based on the optical excitation of neutral atoms to Rydberg states with principal quantum number $n\gg 1$ currently appear as the most promising scheme for the implementation of two-qubit gates in our 2D quantum processing architecture. For Rydberg atoms, the evoked dipole moment mediates strong long-range interaction exceeding the accessible coupling strength of groundstate atoms by orders of magnitude. Central to quantum information processing with Rydberg atoms is the shift of the resonance frequency of one atom (target atom) induced by the excitation of a nearby control atom to a Rydberg state \cite{PhysRevLett.85.2208,PhysRevA.65.052301,PhysRevA.72.022347,RevModPhys.82.2313}. The consequence is a blockade sphere of inhibited excitation with a radius in the micrometer regime which enables quantum information architectures with qubit separations up to $10 \micro\meter$. This has been experimentally demonstrated in systems of two dipole traps \cite{PhysRevLett.104.010502,PhysRevLett.104.010503,PhysRevA.82.030306} with typical trap parameters of $\unit{3.2}{\micro\meter}$ waist and $\unit{8.7}{\micro\meter}$ separation as given in \cite{PhysRevA.82.030306}.
As presented in Fig. \ref{fig:microoptics}(d,e), we currently operate 2D arrays of well resolved traps with a fundamental pitch of \unit{13}{\micro\meter} and a beam waist of \unit{1.6}{\micro\meter}, which shows that by minor modifications in the optical setup we can reach and even exceed the required trap parameters for a successful Rydberg-gate operation as demonstrated in \cite{PhysRevA.82.030306}. For this reason, we can apply to our system the detailed analysis of Rydberg state mediated quantum computing with focus on the relevant physical mechanisms contributing to gate errors given in \cite{PhysRevA.82.030306,PhysRevA.72.022347,1742-6596-264-1-012023}
which predicts a fidelity well above 0.99. The expected intrinsic error evaluates to $6.5\times 10^{-3}$ and the experimentally determined fidelity is 0.92, where technical errors contribute most of the difference between the prediction and the experimental results.\\ This discussion shows that Rydberg-mediated two-qubit gate operations have become a very promising candidate for the currently still open issue of implementing a suitable two-qubit quantum gate in our 2D quantum processing architecture.

\section{Conclusion}
\label{sec:conclusion}

We have presented a scalable architecture for quantum information processing and quantum simulation with neutral atoms and have discussed its characteristics with regard to the requirements imposed for the successful implementation of quantum computing schemes. The design is based on a 2D quantum register created by 2D sets of optical micro-potentials, which can be conceptionally split into spatially separated functional units, e.g. for preparation, processing, and readout.\\
We obtain a suitable hardware with typical dimensions of the individual register cell in the range of a few microns from arrays of focused laser beams employing micro-fabricated lens arrays. This implementation ensures single-site addressability as demonstrated by producing reconfigurable trap patterns and quantum state control of selected qubits. In a combined system used for single-site addressing, the stability and the diffraction-limited performance of the microlens array is complemented by the flexibility of a per-pixel addressable spatial light modulator. The introduced system is capable of performing single-qubit operations as well as qubit specific initialization and readout.\\
A scalable quantum shift register connecting adjacent trapping sites realizes the data bus of the prospect quantum processor. As demonstrated experimentally, the shift operation can be performed with negligible atom loss, heating, or additional dephasing or decoherence, thus allowing for qubit transport over macroscopic distances. Intrinsically, the quantum shift register provides full control over trap separations which also becomes an essential ingredient for the implementation of two-qubit gates. In this respect, among other approaches, the optical and therefore site-selective control of Rydberg interaction turns out to be a very promising candidate for implementing two-qubit gates for quantum computing in 2D quantum registers.\\
In summary, we have given a detailed analysis of our architecture for scalable quantum information processing with neutral atoms in 2D quantum state registers. Although the quantum processor has yet to be implemented in full operation by combining all of the building blocks as they have been analyzed in the previous sections with two-qubit gate operations as achieved in refs. \cite{PhysRevLett.104.010502,PhysRevLett.104.010503,PhysRevA.82.030306}, no principle obstacles can be identified to prevent a successful realization.
In addition, next-generation configurations will strongly benefit from the technological basis available in micro-fabrication which will enable optical-, semiconductor- and micro-mechanical structures to be combined on a single chip and therefore opening an excellent path towards parallelized, large scale quantum computing.

\begin{acknowledgments}
We acknowledge financial support by the Deutsche Forschungsgemeinschaft (DFG), by the European Commission (Integrated Projects ACQUIRE, ACQP, and SCALA), by NIST (Grant No. 60NANB5D120), and by the DAAD (Contract No. 0804149).
\end{acknowledgments}

\bibliographystyle{apsrev4-1}
\bibliography{QUIPS}   




\end{document}